\documentclass[11pt, a4paper]{article}

\usepackage[margin=2cm]{geometry}
\usepackage{cite}
\usepackage{amsmath}
\usepackage{amsfonts}
\usepackage{amssymb}
\usepackage{graphicx}
\usepackage{hyperref}
\usepackage{color}
\usepackage{float}
\usepackage[utf8]{inputenc}
\usepackage[toc,page]{appendix}

\begin{document}

\title{{\bf \Large Horizon geometry for Kerr black holes with synchronised hair}}

\vspace{0.5cm}

 \author{
 {\large Jorge F. M. Delgado}\footnote{jorgedelgado@ua.pt}, \
{\large Carlos A. R. Herdeiro}\footnote{herdeiro@ua.pt} \ and
{\large Eugen Radu}\footnote{eugen.radu@ua.pt} 
\\ 
\\
{\small Departamento de F\'\i sica da Universidade de Aveiro and} \\ 
{\small Center for Research and Development in Mathematics and Applications (CIDMA)} \\ 
{\small   Campus de Santiago, 3810-183 Aveiro, Portugal}
}


\date{April 2018}
\maketitle

\begin{abstract}
We study the horizon geometry of Kerr black holes (BHs) with scalar synchronised hair~\cite{Herdeiro:2014goa}, a family of solutions of the Einstein-Klein-Gordon system that continuously connects to vacuum Kerr BHs.  We identify the region in parameter space wherein a global isometric embedding in Euclidean 3-space, $\mathbb{E}^3$,  is possible for the horizon geometry of the hairy BHs. For the Kerr case, such embedding is possible iff the horizon  dimensionless spin $j_H$ (which equals the total dimensionless spin, $j$), the sphericity $\mathfrak{s}$ and the horizon linear velocity $v_H$ are smaller than critical values, $j^{\rm (S)},\mathfrak{s}^{\rm (S)}, v_H^{\rm (S)}$, respectively. For the hairy BHs, we find that
$j_H<j^{\rm (S)}$ is a sufficient, but not necessary, condition for being embeddable; $v<v_H^{\rm (S)}$ is a necessary, but not sufficient, condition for being embeddable; whereas  $\mathfrak{s}<\mathfrak{s}^{\rm (S)}$ is a necessary and sufficient condition for being embeddable in $\mathbb{E}^3$. Thus the latter quantity provides the most faithful diagnosis for the existence of an $\mathbb{E}^3$ embedding within the whole family of solutions.  We also observe that sufficiently hairy BHs are always embeddable, even if $j$ -- which for hairy BHs (unlike Kerr BHs) differs from $j_H$ --, is larger than unity. 
\end{abstract}

\section{Introduction}
For slow rotation, the spatial sections of the Kerr black hole (BH)~\cite{PhysRevLett.11.237} event horizon\footnote{Hereafter we refer to the geometry of the horizon spatial sections as the ``horizon geometry", for simplicity.} are oblate spheroids. These are closed 2-surfaces, with everywhere positive Gaussian curvature, and with a proper size for the equatorial geodesic circle larger than for meridian ones. This fact is intuitive: rotation typically deforms spherical objects into oblate ones, as it is familiar for the (rotating) Earth. Everywhere positively curved 2-surfaces can be \textit{globally} embedded in Euclidean 3-space, $\mathbb{E}^3$, and the embedding is rigid ($i.e.$ unique up to rigid rotations - see $e.g.$~\cite{Gibbons:2009qe} and references therein). Thus, the horizon geometry of slowly spinning Kerr BHs can be embedded in $\mathbb{E}^3$ and this embedding provides a rigorous and intuitive tool to visualise it -- see~\cite{Smarr:1973zz,Frolov:2006yb,Gibbons:2009qe} and Section~\ref{secemb1} below.

For sufficiently fast rotation, however, the Kerr horizon geometry becomes more exotic.  Smarr~\cite{Smarr:1973zz} first observed that for a dimensionless spin $j > {\sqrt{3}}/{2}\equiv j^{\rm (S)}$ -- hereafter the \textit{Smarr point} --, the Gaussian curvature of the horizon becomes negative in a vicinity of the poles. In this regime, an isometric embedding of the Kerr horizon geometry in $\mathbb{E}^3$ is no longer possible. In fact, such embedding fails even \textit{locally} at the negatively curved axi-symmetry fixed points (the poles). The Kerr horizon may still be embedded in other manifolds, such as Euclidean 4-space~\cite{Frolov:2006yb}, hyperbolic 3-space~\cite{Gibbons:2009qe} or 3-dimensional Minkowski spacetime (see $e.g.$~\cite{Kleihaus:2015aje,Shoom:2015rda}); these do not provide, however, a simple and intuitive visualisation of the Kerr horizon geometry.

In this paper, we shall study the horizon geometry of a generalisation of the Kerr solution and, in particular its embedding in  $\mathbb{E}^3$. The generalised Kerr BHs we shall be considering are Kerr BHs with synchronised scalar hair~\cite{Herdeiro:2014goa} - see also~\cite{Herdeiro:2015gia}. This is a family of stationary, non-singular (on and outside) the event horizon, asymptotically flat BH solutions of Einstein-Klein-Gordon theory, that interpolates between Kerr BHs (the ``bald" limit) and horizonless, everywhere regular boson stars~\cite{Schunck:2003kk} (the solitonic limit). The generic solution possesses an event horizon surrounded by a non-trivial scalar field configuration. It circumvents well known no-hair theorems (see $e.g$~\cite{Bekenstein:1972ny,Herdeiro:2015waa,Sotiriou:2015pka,Volkov:2016ehx}) due to a synchronisation of the angular velocity of the horizon, $\Omega_H$ with the angular phase velocity of the scalar field (see also~\cite{Dias:2011at,Herdeiro:2014ima}). Various other asymptotically flat, four dimensional ``hairy" BHs akin to these solutions have been constructed in~\cite{Kleihaus:2015iea,Herdeiro:2015tia,Herdeiro:2016tmi,Delgado:2016jxq}.

These hairy BHs possess, at the level of their event horizon, only the same two conserved charges of a Kerr BH: the horizon mass, $M_H$ and the horizon angular momentum, $J_H$, that can be computed as Komar integrals~\cite{Komar:1958wp}. It has been previously observed that the corresponding dimensionless spin $j_H\equiv J_H/M^2_H$ can exceed unity~\cite{Herdeiro:2015moa}, unlike that of the vacuum Kerr BH. On the other hand, a geometrically meaningful horizon linear velocity $v_H$ ($cf.$ Section~\ref{secemb1}) never exceeds unity ($i.e.$ the velocity of light), for either vacuum Kerr or the hairy BHs, and it only attains unity  for extremal vacuum Kerr~\cite{Herdeiro:2015moa}.  This suggests that by virtue of the coupling between the horizon and the surrounding ``hair", these BHs present a different ability to sustain angular momentum; heuristically, they have a different (higher) moment of inertia. Such observation makes these hairy BHs an interesting laboratory to test the relation between rotation, angular momentum and horizon deformability.

Here, we shall compare the embedding of these hairy BHs in $\mathbb{E}^3$ with that of the Kerr BH. For the latter, such embedding is only possible if the horizon dimensionless spin $j_H$ (which for the Kerr case equals the total spacetime dimensionless spin, $j$), the sphericity $\mathfrak{s}$ and the horizon linear velocity $v_H$ are smaller than critical values attained at the Smarr point: $j^{\rm (S)},\mathfrak{s}^{\rm (S)}, v_H^{\rm (S)}$. As we shall see, this family of hairy BHs allows us to disentangle the role of the angular momentum and the linear velocity in terms of being compatible with embeddable solutions. Our results indicate that  $j^{\rm (S)}$ ($v_H^{\rm (S)}$) provides only a minimum (maximum) value below (above) which the Euclidean embedding is guaranteed (impossible). On the other hand, it is  the sphericity $\mathfrak{s}^{\rm (S)}$, that remains a faithful diagnosis for the existence of the $\mathbb{E}^3$ embedding, throughout the whole family of solutions, below (above) which the Euclidean embedding is guaranteed (impossible). We shall also observe that sufficiently hairy BHs are always embeddable, even if the \textit{total} dimensionless spin, $j$, is larger than unity. 

This paper is organised as follows. In Section~\ref{section2} we review the isometric embedding of Kerr BHs in $\mathbb{E}^3$, in particular presenting the threshold sphericity and horizon linear velocity. The corresponding analysis for hairy BHs is performed in Section~\ref{section3}, detailing the behaviour of the sphericity, dimensionless spin and horizon linear velocity throughout the parameter space. Concluding remarks are presented in Section~\ref{section4}.

\section{Kerr Black Holes}
\label{section2}

\subsection{Isometric embedding}
\label{secemb1}

The Kerr BH~\cite{PhysRevLett.11.237} is the unique 4-dimensional, regular on and outside the event horizon, axisymmetric and asymptotically flat solution of vacuum Einstein's gravity. In Boyer-Lindquist coordinates~\cite{Boyer:1966qh}, the metric reads:
\begin{equation}
\label{Kerr}
  ds^2 = -\frac{\Delta}{\Sigma} \left( dt - a \sin^2 \theta d\varphi \right)^2 + \Sigma \left( \frac{dr^2}{\Delta} + d\theta^2 \right) + \frac{\sin^2 \theta}{\Sigma} \left[ a dt - \left( r^2 +a^2 \right) d\varphi \right]^2 \hspace{3pt} ,
\end{equation}
where $\Delta \equiv  r^2 - 2Mr + a^2$ and $\Sigma \equiv r^2 + a^2 \cos^2 \theta$, in which, $M$ is the ADM mass and $a \equiv J/M$ is the total angular momentum per unit mass.

The isometric embedding of the Kerr horizon in $\mathbb{E}^3$ has been first considered by Smarr~\cite{Smarr:1973zz}. Let us reconstruct the main result. The 2-metric induced in the spatial sections of the event horizon is obtained as a $t=$constant, $r=r_H$ section of~\eqref{Kerr}, where $r_H\equiv M+\sqrt{M^2-a^2}$ is the largest root of $\Delta=0$. This 2-metric reads: 
\begin{equation}\label{InducedMetricKerr}
  d\sigma^2 = \Sigma_{r_H}\ d\theta^2 + \frac{\sin^2 \theta}{\Sigma_{r_H}} \left( r_H^2 + a^2  \right)^2 d\varphi^2 \hspace{3pt} ,
\end{equation}
where $\Sigma_{r_H} \equiv  r_H^2 + a^2 \cos^2 \theta$. The embedding of any $U(1)$ invariant 2-surface in $\mathbb{E}^3$, with the standard Cartesian metric $d\sigma^2 = dX^2 + dY^2 + dZ^2$, may be attempted by using the following embedding functions,
\begin{eqnarray}
  X+iY = f(\theta) e^{i \varphi}  \ , \qquad   Z = g(\theta) \  ,
\end{eqnarray}
so that,
\begin{equation}\label{InducedMetricFlat3DSpace}
  d\sigma^2 = \left( f'(\theta)^2 + g'(\theta)^2 \right) d\theta^2 + f(\theta)^2 d\varphi^2 \hspace{3pt} ,
\end{equation}
where the prime $'$ indicates the derivative with respect to $\theta$. Comparing with \eqref{InducedMetricKerr}, we obtain, for the Kerr horizon, 
\begin{equation}
  f'(\theta)^2 + g'(\theta)^2 = \Sigma_{r_H} \hspace{10pt} , \hspace{10pt}  f(\theta) = \frac{\sin \theta  \left( r_H^2 + a^2 \right)}{\sqrt{r_H^2 + a^2 \cos^2 \theta}} \hspace{3pt} . \end{equation}
Thus,
\begin{equation}
  g'(\theta) = \frac{\sqrt{h(\theta) }}{\left( r_H^2 + a^2 \cos^2 \theta \right)^{3/2}} \ , \qquad  h(\theta)\equiv (r_H^2 + a^2 \cos^2 \theta)^4 - (r_H^2 + a^2)^4 \cos^2 \theta \ .
  \label{gprime}
\end{equation}

In order for the Kerr horizon to be embeddable, both embedding functions have to be real. In the case of $g(\theta)$ this requires $h(\theta) \geqslant 0$.
Since $h(0)=0=h'(0)$ and $ h''(0) = 2 \left( r_H^2 + a^2 \right)^3 \left( r_H^2 - 3 a^2 \right)$,  then $h''(0) \leqslant 0$ iff  $|a|\geqslant {r_H}/{\sqrt{3}}$. In this region of the parameter space the Kerr horizon fails to be embeddable in $\mathbb{E}^3$ (in the neighbourhood of the poles). In the complementary domain 
\begin{equation}
|a| \leqslant \frac{\sqrt{3}}{2} M \ ,
\end{equation}
the Kerr horizon is embeddable. The threshold of this inequality $|j^{\rm (S)}|=|a| /M= {\sqrt{3}}/{2}$, defines a point for the (absolute value of the) dimensionless spin $j\equiv a/M$,  which we shall call the \textit{Smarr point}.

The Gaussian curvature for~\eqref{InducedMetricKerr}  is 
\begin{equation}
\mathcal{K}=\frac{R}{2}=\frac{\left( r_H^2 + a^2 \right) \left( r_H^2 - 3 a^2\cos^2\theta \right)}{(\Sigma_{r_H})^3} \ ,
\end{equation}
where $R$ is the Ricci scalar.  Thus, at the poles, the Gaussian curvature has the same sign as $h''(0)$. Therefore, the Euclidean embedding fails precisely when the Gaussian curvature at the poles becomes negative.

\subsection{Sphericity  and horizon linear velocity} 
\label{subsec:HLV}
We now introduce one geometrical and one physical parameter related to the horizon: the sphericity, measuring how much the horizon intrinsic shape deviates from a round sphere, and the horizon linear velocity~\cite{Herdeiro:2015moa}, providing a \textit{linear} velocity measure of how fast the null geodesic generators of the horizon are  being dragged with respect to a static observer at spatial infinity. 

We define the \textit{sphericity}, $\mathfrak{s}$, of a $U(1)$ invariant 2-surface as 
\begin{equation}
\mathfrak{s}\equiv \frac{L_e}{L_p} \ ,
\end{equation} 
where $L_e$ is the equatorial proper length and $L_p$ is twice  the proper length of a meridian ($\varphi=$constant curve for an azimuthal coordinate adapted to the $U(1)$ symmetry). Observe that the $U(1)$ symmetry guarantees all meridians have the same proper size. The Kerr horizon has a $\mathbb{Z}_2$ symmetry and the equator is the set of fixed points of this symmetry. Then, from \eqref{InducedMetricKerr}, 
\begin{equation}
  L_e =  \int_0^{2\pi} d\varphi\ \frac{r_H^2 + a^2}{r_H} = 2\pi \frac{r_H^2 + a^2}{r_H} \ , \qquad 
  L_p  = 2 \int_0^\pi d\theta\ \sqrt{r_H^2 + a^2 \cos^2 \theta} = 4 \sqrt{r_H^2 + a^2}\ \text{E} \left( \frac{a^2}{r_H^2 + a^2} \right)  \hspace{3pt},
\end{equation} 
where E($k$) is the complete elliptic integral of second kind: $E(k) \equiv \int_0^{\pi/2} d\theta \sqrt{1 - k \sin^2 \theta}$ \cite{abramowitz1964handbook}. Thus
\begin{equation}\label{eq:ratioKerr}
 \mathfrak{s}(j)= \frac{\pi}{2}  \frac{\sqrt{2+2\sqrt{1-j^2}}}{1+\sqrt{1-j^2}} \frac{1}{E \left( \frac{j^2}{2+2\sqrt{1-j^2}} \right)} \ , \qquad \Rightarrow \qquad \mathfrak{s}^{\rm (S)}\equiv \mathfrak{s}(j^{\rm (S)}) = \frac{\pi}{\sqrt{3}} \frac{1}{E(1/4)} \approx 1.23601\ .
\end{equation}

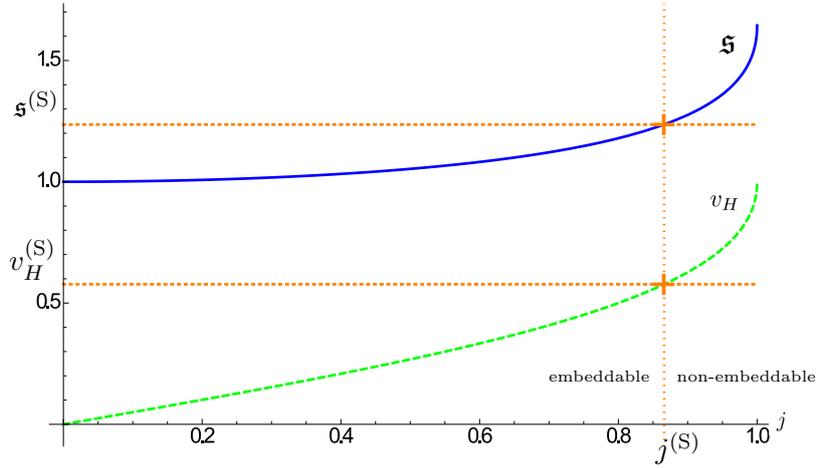
\begin{figure}
  \centering
  \includegraphics[scale=0.70]{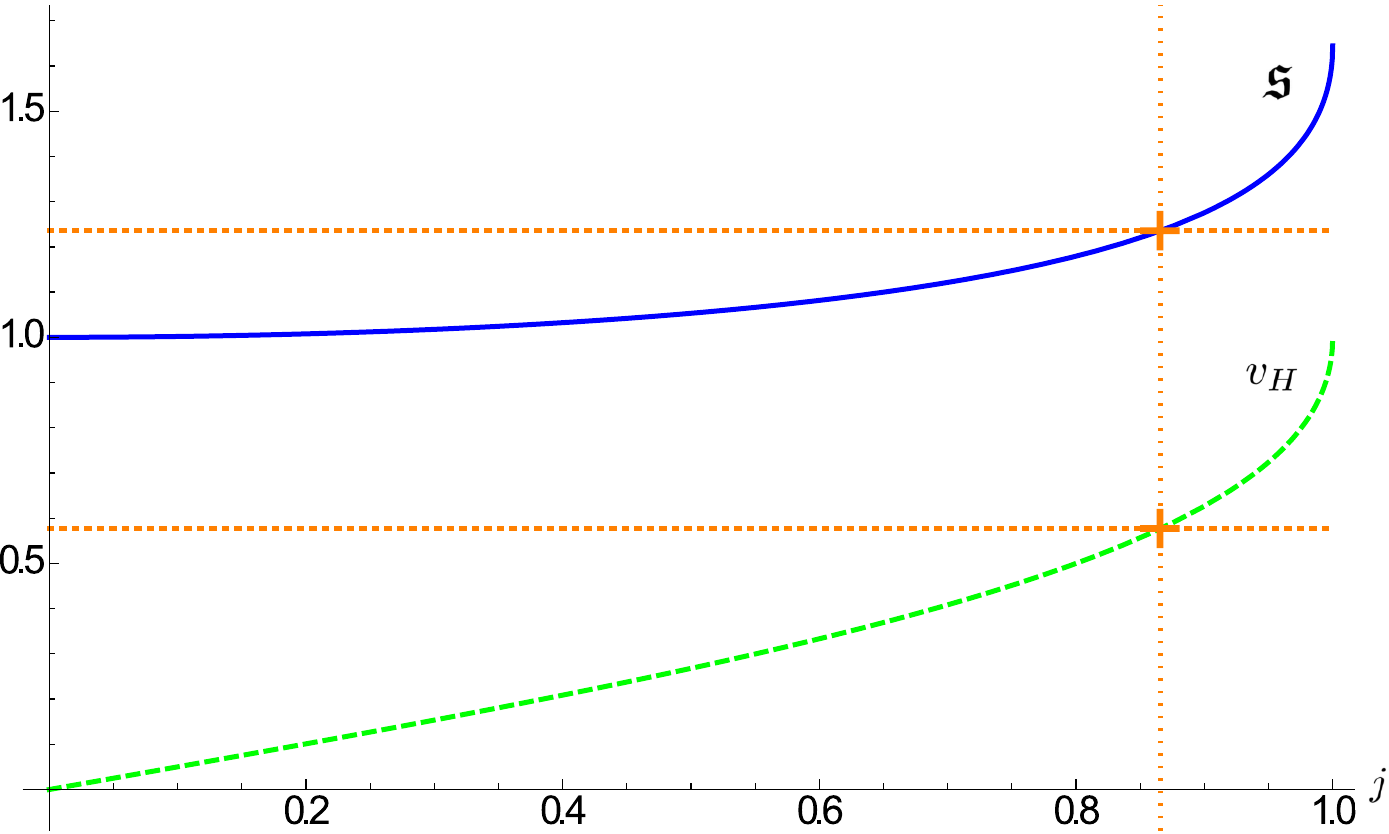}
\begin{picture}(0,0)
\put(-293,125){$\mathfrak{s}^{\rm (S)}$}
\put(-295,67){$v_H^{\rm (S)}$}
\put(-53,-4){$j^{\rm (S)}$}
\put(-93,25){\tiny{embeddable}}
\put(-45,25){\tiny{non-embeddable}}
\end{picture}

  \caption{The sphericity $\mathfrak{s}$ (top solid blue curve) and the horizon linear velocity, $v_H$ (bottom dashed green curve) as a function of the dimensionless angular momentum, $j$, for the Kerr horizon. The vertical line gives the Smarr point dimensionless spin value, $j^{\rm (S)}$.}
  \label{fig:RatioLeLpKerr}
\end{figure}

In Fig. \ref{fig:RatioLeLpKerr} we exhibit the sphericity for the Kerr horizon as function of the dimensionless angular momentum, $j$. One can see that the sphericity is always $\mathfrak{s}\geqslant 1$, meaning that the length along the poles is always equal or smaller that the length along the equator. Thus, the Kerr event horizon can be a round sphere (for $\mathfrak{s}= 1$, corresponding to the Schwarzschild limit) or an oblate spheroid (for $\mathfrak{s} > 1$). In the figure we highlight the Smarr point $j^{\rm (S)}$; the corresponding sphericity, yielding the maximal value of this quantity for which the Kerr horizon is embeddable in $\mathbb{E}^3$, is $\mathfrak{s}^{\rm (S)}$.

The \textit{horizon linear velocity} is defined as~\cite{Herdeiro:2015moa}:
\begin{equation}
v_H \equiv R \Omega_H \ ,
\end{equation}
where $\Omega_H$ is the horizon angular velocity and $R \equiv L_e/2\pi$ is   the circumference radius of the equator. For the specific case of the Kerr BH~\cite{Herdeiro:2015moa,Delgado:2016zxv}, 
\begin{equation}
v_H(j) = \frac{j}{1+\sqrt{1-j^2}} \ , \qquad   \Rightarrow  \qquad   v_H^{\rm (S)}\equiv v_H(j^{\rm (S)}) = \frac{1}{\sqrt{3}} \approx 0.57735  \ .
\end{equation}
In Fig. \ref{fig:RatioLeLpKerr}  we also exhibit  the horizon linear velocity, $v_H$, as a function of the dimensionless angular momentum, $j$, for the Kerr horizon. This velocity never exceeds the speed of light, $v=1$, reaching this value in the extremal case. The horizon linear velocity for the Smarr point, $v_H^{\rm (S)}$, is highlighted in the plot.

\section{Kerr BHs with Scalar Hair}
\label{section3}

Kerr BHs with scalar hair (KBHsSH) \cite{Herdeiro:2014goa} are 4-dimensional, regular on and outside the event horizon, axisymmetric, asymptotically flat solutions of the Einstein-Klein-Gordon action,
\begin{equation}
  \mathcal{S} = \int d^4 x \sqrt{-g} \left[ \frac{1}{16\pi} R - \frac{1}{2} g^{\alpha\beta} \left( \partial_\alpha \Psi^* \partial_\beta \Psi + \partial_\alpha \Psi \partial_\beta \Psi^* \right) - \mu^2 \Psi^* \Psi \right] \hspace{3pt} ,
\end{equation}
where $\Psi$ is a complex scalar field with mass $\mu$. These solutions are only known numerically~\cite{Herdeiro:2015gia} and have been constructed by using the following \textit{ansatz} for the metric and the scalar field,
\begin{eqnarray}
\label{eq:ansatz}
  &ds^2 = - e^{2F_0(r,\theta)} \left(1 - \frac{r_H}{r}\right) dt^2+ e^{2F_1(r,\theta)} \left( \frac{dr^2}{1 - \frac{r_H}{r}} + r^2 d\theta^2 \right) + e^{2F_2(r,\theta)} r^2 \sin^2 \theta \left( d\varphi - W(r,\theta) dt \right)^2 \ ,\ \ \   \\
  &\Psi = \phi(r,\theta)\ e^{i(m\varphi - \omega t)} \ ,
\end{eqnarray}
where the functions $F_0$, $F_1$ $F_2$, $W$ and $\phi$ only depend on the $(r,\theta)$ coordinates, $r_H$ is the radial coordinate of the event horizon\footnote{In the Kerr limit, $r_H$ is related to the radial coordinate in Boyer-Lindquist coordinates, $R_H = M + \sqrt{M^2 - a^2}$,  by $r_H = R_H - {a^2}/{R_H}$~\cite{Herdeiro:2015gia}.}, $m \in \mathbb{Z}$ is the azimuthal harmonic index, and $\omega$ is the scalar field frequency.

This family of solutions depends on three physical quantities: the ADM mass, $M$, the total angular momentum, $J$ and a Noether charge that encodes information about the scalar hair, $Q$, but these are not independent (it is a 2D parameter space). Of the discrete set of families labelled by $m$, here we shall focus on solutions with $m=1$.

In Fig. \ref{fig:ParameterSpaceKBHsSH} (left panel) we exhibit the domain of existence of the solutions in an $M$ \textit{versus} $\Omega_H$ diagram. This domain is bounded by three lines:
\begin{figure}[h!]
  \centering
  \includegraphics[scale=0.66]{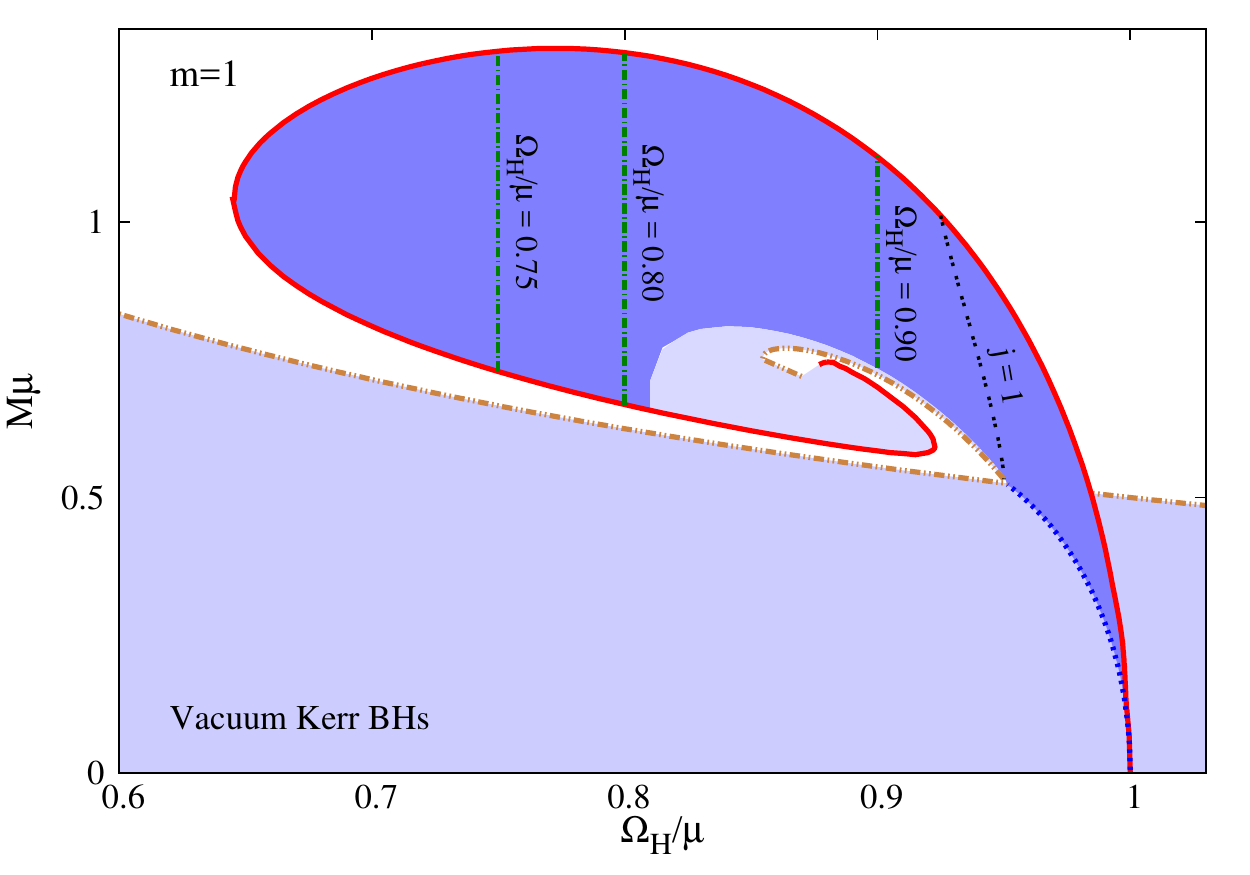}
   \includegraphics[scale=0.66]{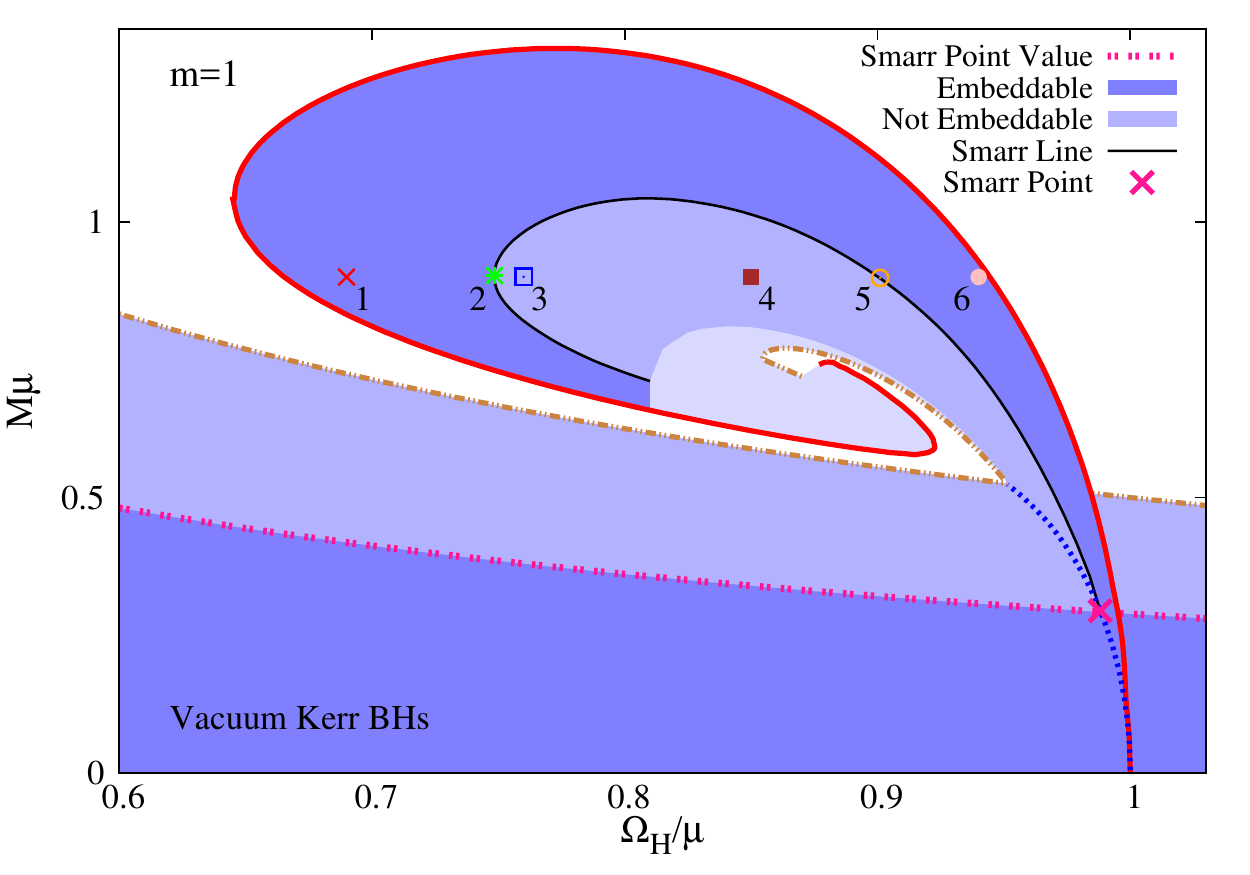}
  \caption{(Left panel) Domain of existence of KBHsSH solutions with $m=1$ (dark blue and upper light blue shaded regions) in an $M$  \textit{versus} $\Omega_H$ diagram, both in units of the scalar field mass $\mu$.  Three lines of constant $\Omega_H$ solutions are displayed; these sequences will be used below. Also the line of solutions with $j=1$ is shown, which starts at the extremal vacuum Kerr. The bottom light blue shaded region corresponds to vacuum Kerr BHs which is limited by a dashed-dotted brown line corresponding to extremal Kerr BHs. (Right panel) Same plot showing the embeddable (dark blue) and not embeddable (intermediate blue) regions of KBHsSH and Kerr BHs delimited by the Smarr line (solid black line) and the Smarr point (pink cross -- extended as the pink dotted line). The lightest blue region of the domain has not been examined in detail (being a numerically challenging region). Six specific solutions are highlighted, whose embedding diagrams are shown in Fig.~\ref{3dembed}. The remaining lines are the same lines as in the left panel.}
  \label{fig:ParameterSpaceKBHsSH}
\end{figure}
\begin{itemize}
  \item The boson star line (red solid line). This is the everywhere regular solitonic limit, entirely composed of a self-gravitating scalar field,  where both the event horizon radius and area goes to zero, $r_H \rightarrow 0$ and $A_H \rightarrow 0$. 
  
  \item The extreme KBHsSH line (dashed-dotted  brown line). These are solutions corresponding to a vanishing Hawking temperature. Due to the choice of coordinates, they also correspond to $r_H \rightarrow 0$, but have a non-zero horizon area, $A_H \nrightarrow 0$, in contrast to the above case.

  \item The existence line (dotted blue line)~\cite{Hod:2012px,Hod:2013zza,Herdeiro:2014goa,Hod:2014baa,Benone:2014ssa,Hod:2016lgi}. This is the vacuum Kerr BH limit (no scalar field). Kerr BHs in this line have $j$ varying from 0 (in the $M\mu\rightarrow 0$ limit) and unity (in the extremal limit).
\end{itemize}

The boson star line and the extremal KBHsSH line are expected to spiral toward a common (likely singular) point, which however has not yet been reached numerically. It proves convenient to introduce two parameters that quantify the amount of ``hair" in the BH:
\begin{equation}
p\equiv \frac{M_\Psi}{M}=1-\frac{M_H}{M} \ , \qquad q\equiv \frac{J_\Psi}{J}=1-\frac{J_H}{J}  \ ,
\end{equation}
where $M_H,J_H$ ($M_\Psi,J_\Psi$) are the horizon (scalar field) mass and angular momentum, that can be computed as Komar integrals on the horizon (volume integrals outside the horizon) - see, $e.g.$~\cite{Herdeiro:2015gia}.

\subsection{Embedding}

To embed the spatial sections of the event horizon of KBHsSH in $\mathbb{E}^3$ we repeat the procedure of Section~\ref{secemb1}. The induced metric of these spatial sections is obtained by setting $t=$constant and $r=r_H$ in~\eqref{eq:ansatz}:
\begin{equation}\label{InducedMetricKBHsSH}
  d\sigma^2 =  r_H^2\left[e^{2F_1(r_H,\theta)} d\theta^2 + e^{2F_2(r_H,\theta)}  \sin^2 \theta d\varphi^2\right] \hspace{3pt}.
\end{equation}
Comparison with eq.~\eqref{InducedMetricFlat3DSpace} yields:
\begin{equation}
  f'(\theta)^2 + g'(\theta)^2 = e^{2F_1(r_H,\theta)} r_H^2  \hspace{10pt} , \qquad   f(\theta) = e^{F_2(r_H,\theta)} r_H \sin \theta \hspace{3pt},
\end{equation}
where the functions $F_1$ and $F_2$ now only depend on $\theta$. It follows that:
\begin{equation}
  g'(\theta) = r_H\sqrt{k(\theta) } \ , \qquad k(\theta)\equiv e^{2F_1(r_H,\theta)} - e^{2F_2(r_H,\theta)} \left[ F'_2(r_H,\theta) \sin \theta + \cos \theta \right]^2 \ .
\end{equation}
As for the Kerr case, the spatial sections of the hairy BHs horizon will only be globally embeddable in $\mathbb{E}^3$ if both embedding functions are real for all $\theta$, thus iff $k(\theta)\geqslant 0$.
Observe that $k(0) = e^{2F_1(r_H,0)} - e^{2F_2(r_H,0)}$. In order to avoid conical singularities, $F_1(r,0)=F_2(r,0)$~\cite{Herdeiro:2015gia}. Thus $k(0) = 0$, just as for the Kerr case. Then, $ k'(0) = 2F_1'(r_H,0) e^{2F_1(r_H,0)} - 4 F_2'(r_H,0) e^{2F_2(r_H,0)}$. For regularity, $F_1'(r_H,0) = 0$ and $F_2'(r_H,0) = 0$ \cite{Herdeiro:2015gia}. Thus $k'(0) = 0$, also, as for the Kerr case. Finally, $ k''(0) = 2 e^{2F_2(r_H,0)} \mathcal{T}$, where $ \mathcal{T} \equiv F_1''(r_H,0) - 3 F_2''(r_H,0) + 1 $, and thus the embedding fails if $\mathcal{T} < 0$.

Since the analogous condition to $k(\theta)\geqslant 0$ fails for Kerr BHs with $j>j^{\rm (S)}$, and since Kerr BHs with $j\in [0,1]$ occur at one of the boundaries of the domain of existence of KBHsSH, there will be a non-embeddable region in the domain of existence of KBHsSH. This region was obtained by analysing the above condition for the numerical data, and it is exhibited as the intermediate blue shaded region in  Fig.~\ref{fig:ParameterSpaceKBHsSH} (right panel). It is bounded by a (black solid) line -- dubbed \textit{Smarr line} -- that  terminates at the Smarr point, in the Kerr limit. The Smarr line seems to  spiral in a similar fashion to the boson star line. 

The failure of the embedding of~\eqref{InducedMetricKBHsSH} in $\mathbb{E}^3$ is again accompanied by the development of a region of negative Gaussian curvature at the poles. To see this, note that the Gaussian curvature for~\eqref{InducedMetricKBHsSH}   is 
\begin{equation}
\mathcal{K}=\frac{e^{-2 {F_1}(r_H,\theta )}}{r_H^2} \left\{1+{F_1}'(r_H,\theta ) \left[{F_2}'(r_H,\theta )+\cot \theta \right]-{F_2}'(r_H,\theta )\left[{F_2}'(r_H,\theta )+2
   \cot \theta \right] -{F_2}''(r_H,\theta )\right\} \ .
\end{equation}
At the poles, $\theta=\{0,\pi\}$,
\begin{equation}
 \mathcal{K}|_{\rm poles} = \frac{e^{-2F_1(r_H,0)}}{r_H^2}\ \mathcal{T} \hspace{3pt} .
\end{equation}
Thus we conclude that, just as for the Kerr case, the embedding fails when the Gaussian curvature becomes negative at the poles.

\begin{figure}
 \centering
 \includegraphics[scale=0.19]{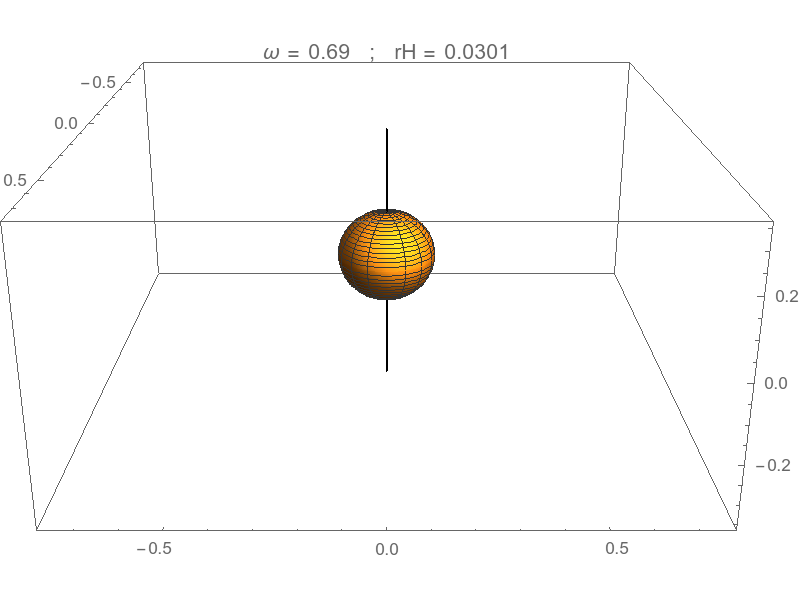}
  \includegraphics[scale=0.19]{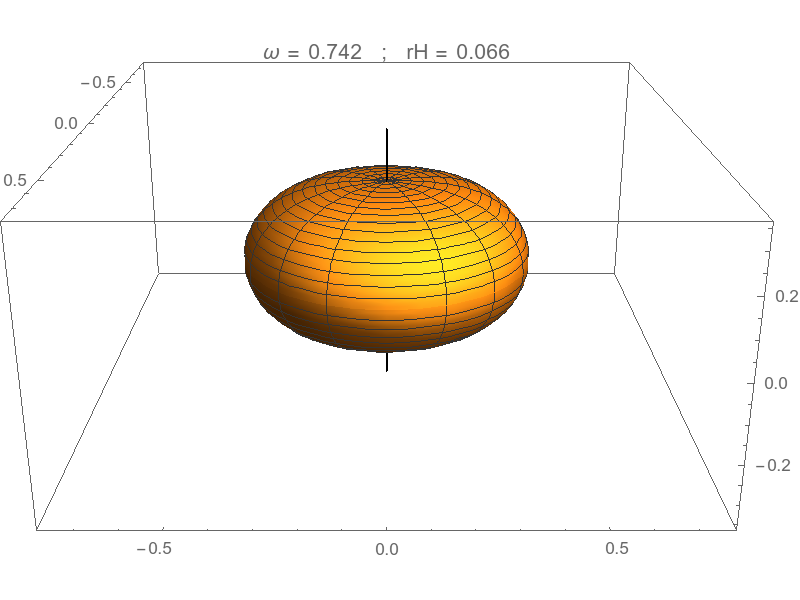}
   \includegraphics[scale=0.19]{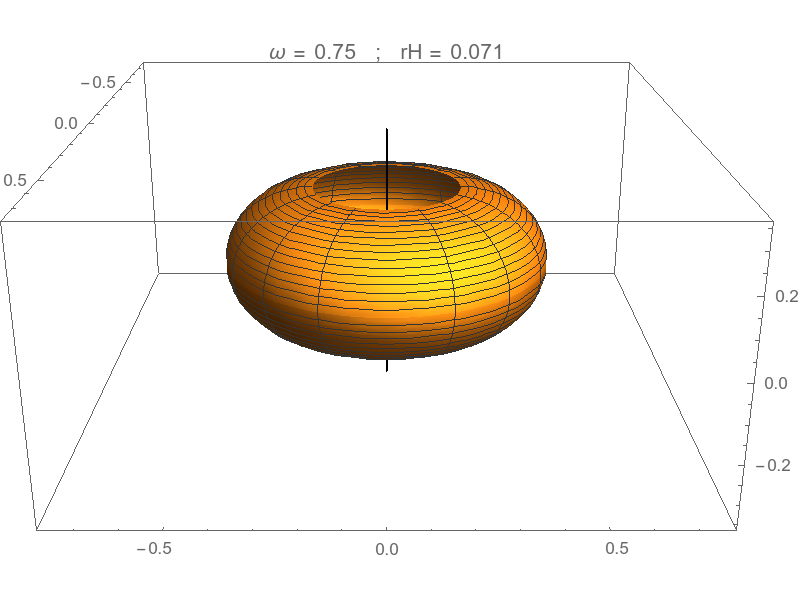}
    \includegraphics[scale=0.19]{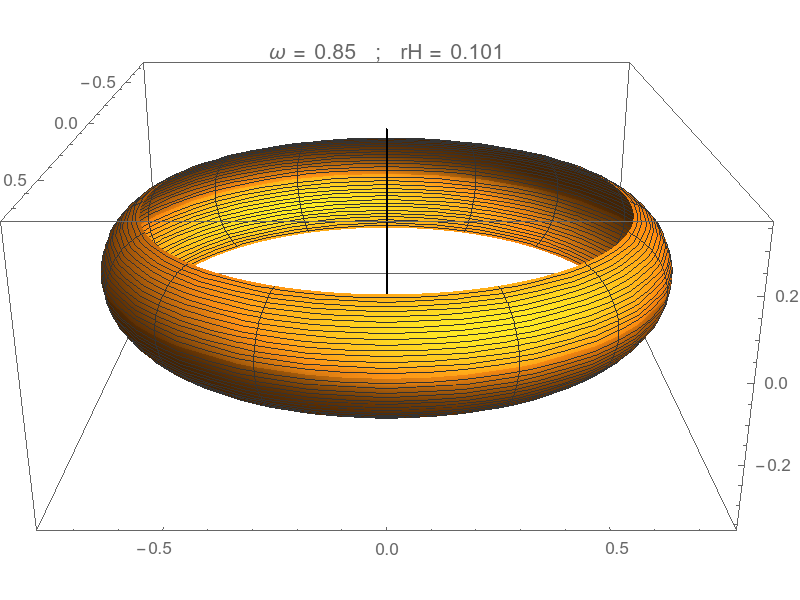}
  \includegraphics[scale=0.19]{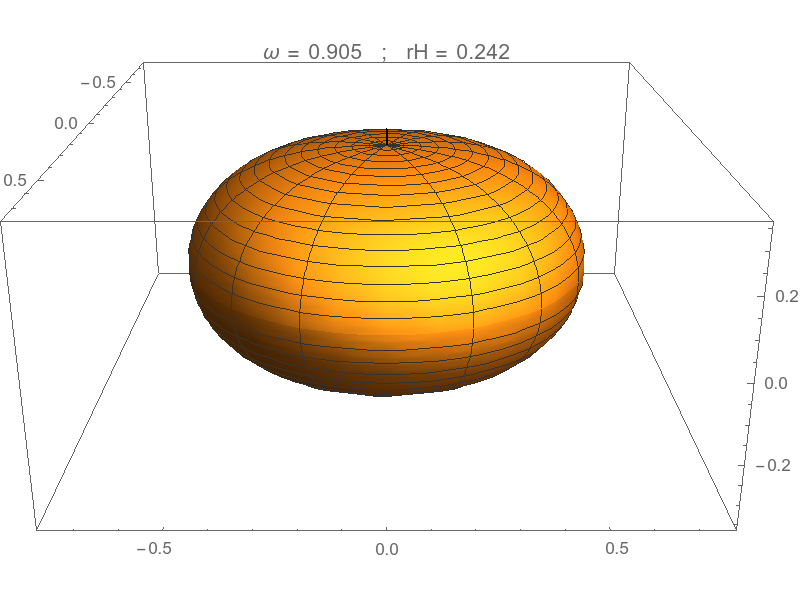}
   \includegraphics[scale=0.19]{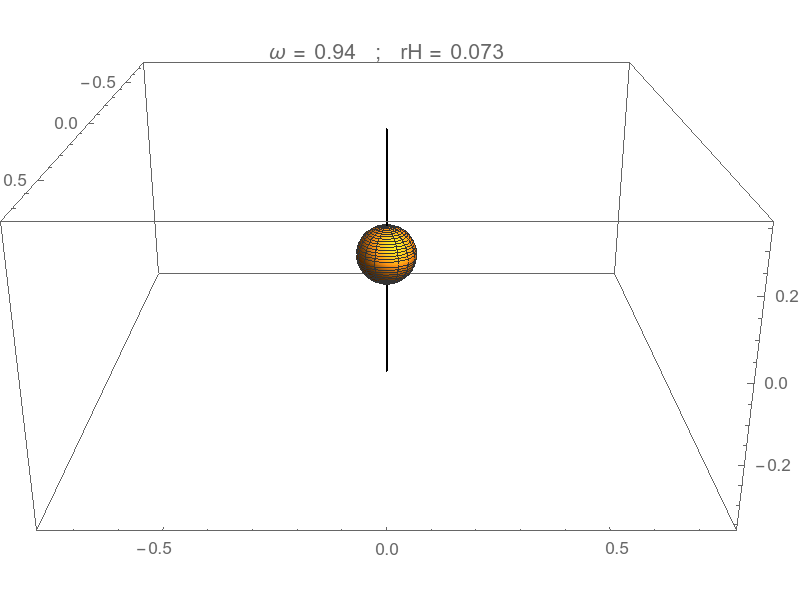}
   \caption{Embedding diagrams,  in $\mathbb{E}^3$, for the horizon of the six highlighted solutions in the right panel of Fig.~\ref{fig:ParameterSpaceKBHsSH}. First (second) row, left to right: solutions 1,2,3 (4,5,6). Solutions 1-6 have parameters $(j;j_H;\mathfrak{s};v_H)$, respectively:  (0.92; 6.27; 1.04; 0.09); (0.88; 5.53; 1.24; 0.29); (0.88; 4.92; 1.27; 0.33); (0.86; 1.69; 1.46; 0.65); (0.89; 1.16; 1.23; 0.48); (1.08; 1.64; 1.01; 0.08).}
  \label{3dembed}
\end{figure}

The 3-dimensional embedding diagrams for the sequence of six solutions highlighted in Fig.~\ref{fig:ParameterSpaceKBHsSH} are shown in Fig.~\ref{3dembed}. These six solutions all have the same $\mu M=0.9$. Solution 1 is in the embeddable region near the boson star line; solution 2 is at the border with the non-embeddable region, wherein solutions 3 and 4 are located. Solution 5 is again at the boundary of the embeddable region and solution 6 is again in the embeddable region close to the boson star line.  Observe that the horizon of the solutions close to the boson star line is quite spherical. The solutions at the threshold of the embeddable region are quite flat near the poles.

Observe that in the representation used in Fig.~\ref{fig:ParameterSpaceKBHsSH}, a $j=$constant vacuum Kerr solution is actually a line in the lower part of the diagram. Thus, the Smarr point of vacuum Kerr BHs is also represented as a line (dashed pink line in the domain of Kerr BHs), separating the globally embeddable region (lower part - dark blue), from the non-embeddable region (upper part - intermediate blue).

\subsection{Sphericity}

For KBHsSH, $L_e$ and $L_p$ can be obtained from the induced metric~\eqref{InducedMetricKBHsSH}, giving:
\begin{equation}
L_e = \int_0^{2\pi} d\varphi\ e^{F_2(r_H,\pi/2)} r_H = 2\pi e^{F_2(r_H,\pi/2)} r_H \ , \qquad 
  L_p  = 2 r_H \int_0^\pi d\theta\ e^{F_1(r_H,\theta)} \ .
\end{equation}
Thus, the sphericity can be written as,
\begin{equation}
  \mathfrak{s}\equiv \frac{L_e}{L_p} = \frac{\pi\ e^{F_2(r_H,\pi/2)}}{\int_0^\pi d\theta\ e^{F_1(r_H,\theta)}} \hspace{3pt}.
\end{equation}

\begin{figure}
  \centering
  \includegraphics[scale=0.8]{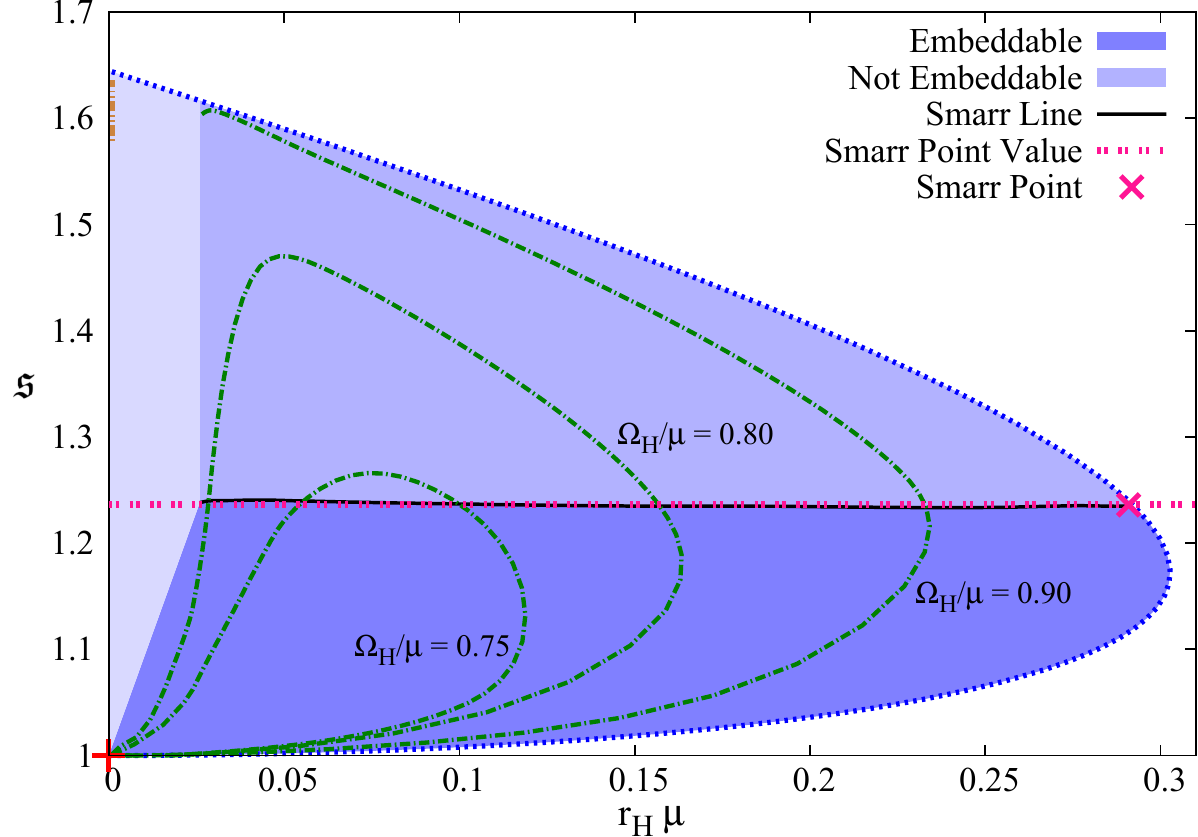}
  \caption{Domain of existence of KBHsSH solutions in a $\mathfrak{s}$  \textit{versus} $r_H$ diagram. As before, the dark blue region corresponds to embeddable solutions, the intermediate blue region corresponds to non-embeddable solutions and the light blue region to the region we have not analysed.  Three lines of constant horizon angular velocity $\Omega_H /\mu = \{ 0.75; 0.80; 0.90 \}$ (dark green dot-dashed lines) are also exhibited. The remaining lines/points are the same as in Fig. \ref{fig:ParameterSpaceKBHsSH}.}
  \label{fig:ratio_rH_KBHsSH}
\end{figure}

In Fig. \ref{fig:ratio_rH_KBHsSH} we exhibit the domain of existence of KBHsSH in a  $\mathfrak{s}$ \textit{versus} $r_H$ diagram. In such a diagram, all boson star solutions and extremal hairy BHs must lie along the $y$ axis, since, as observed above, in both these cases $r_H\rightarrow 0$. It turns out that all  boson star solutions fall into a single point, precisely at the origin of this diagram (red cross), corresponding to $r_H = 0$ and $\mathfrak{s}= 1$. Since boson stars have no horizon, the sphericity here is defined by continuity, and this result means that the event horizon of a KBHsSH very close to the boson star limit will be essentially spherical. This observation resonates with the idea~\cite{Herdeiro:2009qy,Herdeiro:2015moa} that the heavy scalar environment around the hairy BH horizon endows the horizon with a large momentum of inertia. Dragging such a ``heavy'' environment results in a slower horizon linear velocity, as will be confirmed in the next subsection, and therefore in a more spherical event horizon, regardless of the $j$ parameter of the spacetime, which can be even larger than unity. 

All the extremal KBHsSH solutions fall onto a line at $r_H=0$ and $\mathfrak{s} \in [1, 1.64473]$. The largest value in this interval correspond to extremal vacuum Kerr:
\begin{equation}
  \left. \mathfrak{s} (j=1)\right|_{\text{Kerr}} = \frac{\pi}{\sqrt{2}} \frac{1}{E(1/2)} \approx 1.64473 \hspace{3pt} .
\end{equation}
 The region closest to the lowest value in this interval ($\mathfrak{s} = 1$) was not obtained with our numerical solutions, since these solutions correspond to the very central region of the spiral, and hence numerically challenging.  The solutions we have actually examined are a subset of these and are represented as a brown dashed-dotted line in Fig. \ref{fig:ratio_rH_KBHsSH}. 

Fig. \ref{fig:ratio_rH_KBHsSH} also shows the Smarr (black solid) line, as well the Smarr point for Kerr BHs (pink cross), which, for reference, is extended for other values of $r_H$ as a pink dotted line. One can see that the Smarr line has the same value of spheroidicity as the Smarr point, for all solutions analysed, within numerical error. The green dot-dash lines in Fig. \ref{fig:ratio_rH_KBHsSH} correspond to solutions with a constant horizon angular velocity $\Omega_H/\mu = \{ 0.75; 0.80; 0.90 \}$. For each value of $\Omega_H/\mu$ there are two possible values of $r_H$ for the same sphericity. These solutions can be quite different but they are either both embeddable or both non-embeddable, since this property is determined by the value of $\mathfrak{s}$.

\subsection{Horizon linear velocity}
Using the same reasoning as in the Kerr case, we obtain
\begin{equation}
  v_H = e^{F_2(r_H,\pi/2)} r_H \Omega_H \hspace{3pt} .
\end{equation}

\begin{figure}[h!]
  \centering
  \includegraphics[scale=0.8]{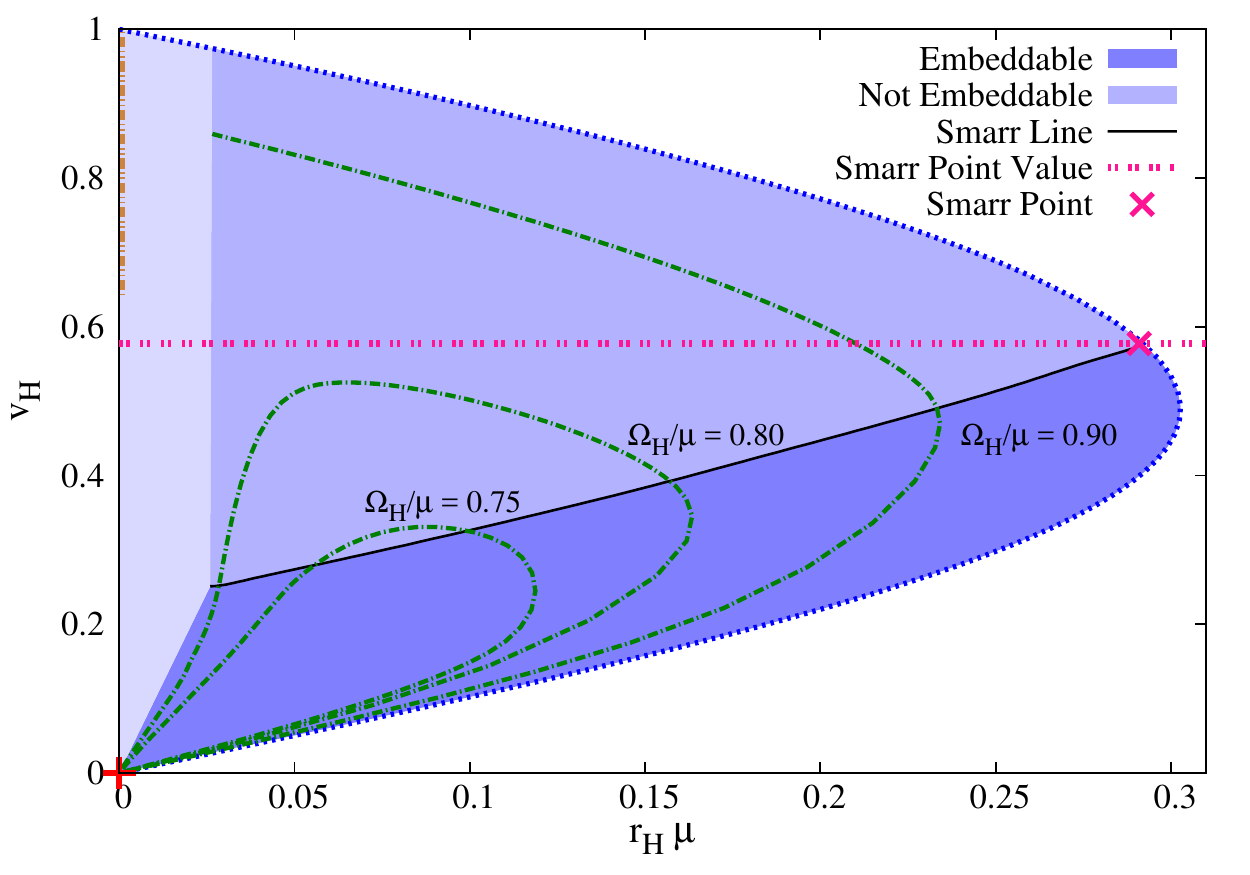}
  \caption{Domain of existence of KBHsSH solutions in a $v_H$  \textit{versus} $r_H$ diagram. All colour coding is as in Fig. \ref{fig:ratio_rH_KBHsSH}.}
  \label{fig:vH_rH_KBHsSH}
\end{figure}

In Fig. \ref{fig:vH_rH_KBHsSH} we exhibit the domain of existence of KBHsSH solutions in a $v_H$ \textit{versus} $r_H$ diagram. 
All boson stars solutions are, in this diagram, a single point -- the origin -- represented as a red cross. 
The remaining points with $r_H=0$ correspond to the extremal hairy BHs. These extremal BHs span the interval $v_H \in [0, 1]$, where $v_H=1$ corresponds to the Kerr limit and $v_H=0$ to the conjectured singular solution at the centre of the spiral.\footnote{Again, our numerical solutions do not extend all the way to $v_H = 0$, for the same reason discussed before.}

The green dot-dashed lines, as before, represent solutions with constant horizon angular velocities, $\Omega_H/\mu = \{ 0.75; 0.80; 0.90 \}$. Taking, for instance, $\Omega_H/\mu = 0.8$, we see that there is a region of degeneracy where it is possible to have two solutions with the same $v_H$,  one being embeddable and the other non-embeddable; $r_H$ or $\mu M$ raise the degeneracy.

Fig. \ref{fig:vH_rH_KBHsSH}  also shows the Smarr (black solid) line, as well the Smarr point for Kerr BHs (pink cross), again extended for other values of $r_H$ as a pink dotted line. In contrast to the case for the sphericity, the Smarr line only matches the value at the Smarr point in the Kerr limit, where $v_H = 1/\sqrt{3}$ -- \textit{cf.} Sec. \ref{subsec:HLV}. 
Here the Smarr line decreases monotonically as $r_H$ decreases. This line is always (for all solutions studied) smaller than the Smarr point value for Kerr BHs.
This indicates that $v_H=1/\sqrt{3}$ is the maximum velocity at which a Kerr BH, with or without scalar hair, can rotate and be globally embeddable in Euclidean 3-space.

\subsection{Horizon angular momentum}
Let us also analyse the relation between the angular momentum carried by the BH and the existence of an Euclidean embedding. In the Kerr case, it is irrelevant if one considers this angular momentum to be the horizon one or the asymptotically measured one, since they coincide. But for the case of the hairy BHs this is not the case, as a part of the angular momentum is stored in the scalar field outside the horizon (similarly to what occurs for the energy). Let us consider the horizon dimensionless angular momentum, $j_H$, as this is really the angular momentum stored in the BH. 

\begin{figure}
  \centering
  \includegraphics[scale=0.66]{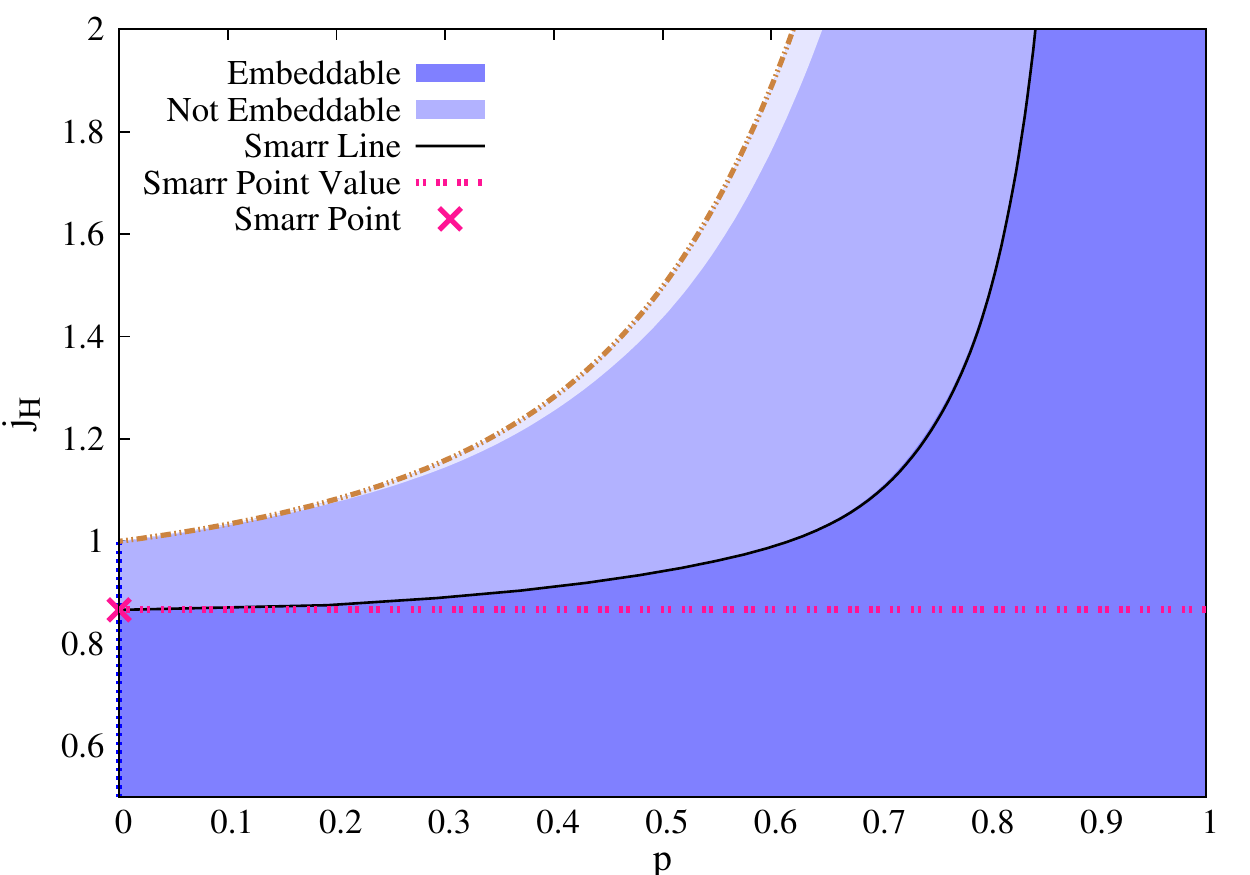}
   \includegraphics[scale=0.66]{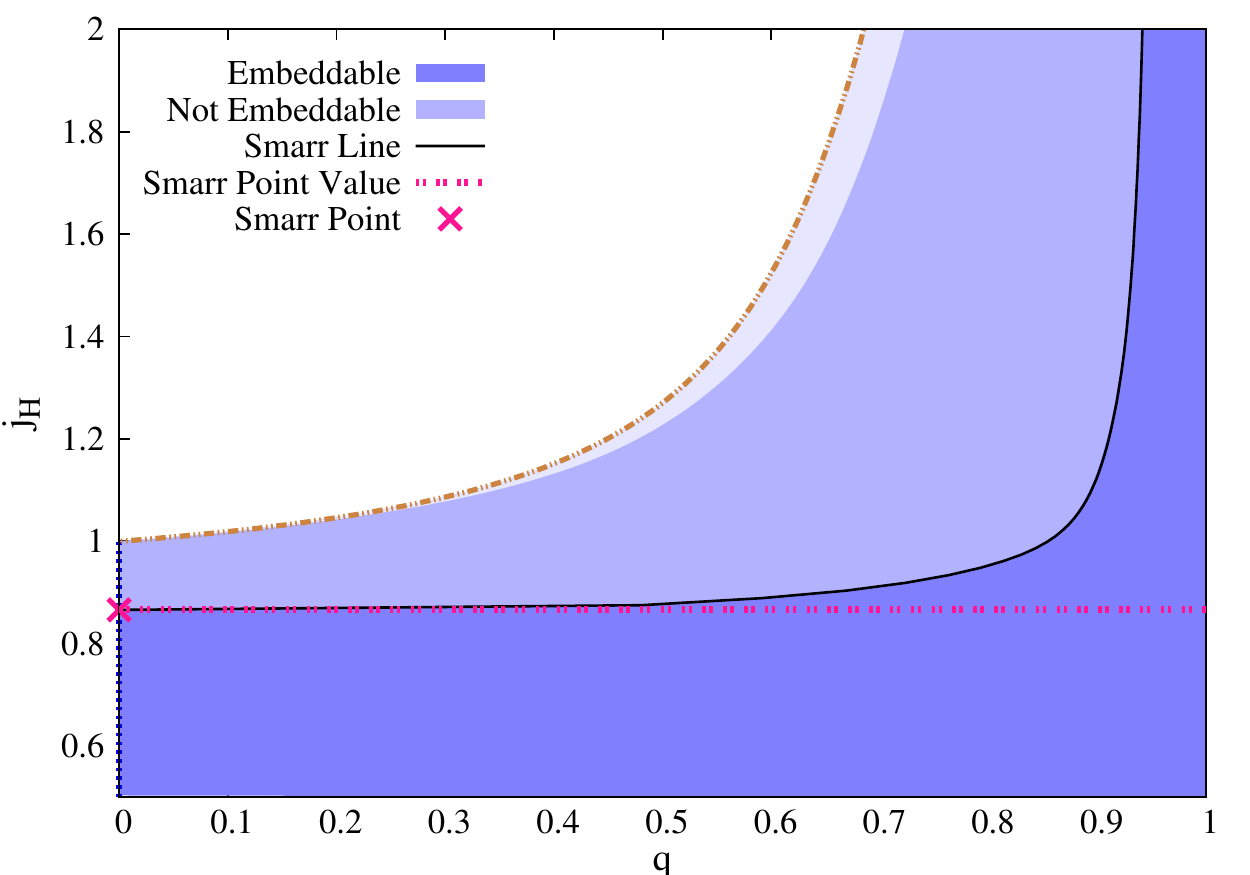}
  \caption{Horizon dimensionless angular momentum, $j_H$ \textit{versus} the hairiness parameters $p$ (left panel) and $q$ (right panel). The color coding is the same as in previous figures.}
  \label{fig:jH}
\end{figure}

In Fig.~\ref{fig:jH}, we present the horizon angular momentum as a function of the hairiness parameters $p$ and $q$. Taking $p=0=q$ is the Kerr limit where we see the separation between embeddable and non-embeddable solutions at the Smarr point. The value $p=1=q$ is the boson star limit where we see all solutions become embeddable. Most importantly, all solutions with $j_H<j^{\rm (S)}$ are embeddable. The Smarr dimensionless spin provides a lower bound below which all hairy BHs can be embedded in Eucliden 3-space.  

Concerning the total angular momentum, we would like to point out that in the boson star limit it can become larger than unity. This occurs for the boson stars between the Newtonian limit (at $\Omega_H/\mu=1$) and the $j=1$ point - see Fig.~\ref{fig:ParameterSpaceKBHsSH} (left panel). In the vicinity of these solutions, KBHsSH also have $j>1$ by continuity, and they are always embeddable in $\mathbb{E}^3$, as one may verify from inspection of Fig.~\ref{fig:ParameterSpaceKBHsSH} (right panel).

\section{Conclusions}
\label{section4}

In this paper we have analysed the horizon geometry of Kerr BHs with scalar hair, a family of solutions that continuously connects to vacuum Kerr BHs. We have been particularly interested in distinguishing solutions which are embeddable in Euclidean 3-space, $\mathbb{E}^3$, as this embedding provides an accurate and intuitive tool to perceive the horizon geometry of these BHs. In the Kerr case, the solutions stop being embeddable at the Smarr point, when the dimensionless spin $j$, the sphericity $\mathfrak{s}$ and the horizon linear velocity $v_H$ are smaller than critical values attained at the Smarr point: $j^{\rm (S)},\mathfrak{s}^{\rm (S)}, v_H^{\rm (S)}$. Our analysis of the hairy BHs shows that:
\begin{description}
\item[$\bullet$]  all hairy solutions with a \textit{horizon} dimensionless spin $j_H\leqslant j^{\rm (S)}$ are embeddable in $\mathbb{E}^3$;  but there are also hairy BHs with $j_H> j^{\rm (S)}$ that are embeddable;
\item[$\bullet$]  $\mathfrak{s}^{\rm (S)}$ remains the threshold sphericity for the embeddable solutions all over the explored domain of existence of hairy BHs. Thus hairy BHs with  $\mathfrak{s}<\mathfrak{s}^{\rm (S)}$ ($\mathfrak{s}>\mathfrak{s}^{\rm (S)}$) are (are not) embeddable;
\item[$\bullet$]  there are hairy solutions with $v_H< v_H^{\rm (S)}$ which are not globally embeddable in $\mathbb{E}^3$ and all hairy BHs with $v_H> v_H^{\rm (S)}$ are not embeddable. 
\end{description}
The analysis of this family of hairy solutions suggestively disentangles the role of these three parameters in the embedding properties of the solutions. For the general family of hairy BHs,  $j_H<j^{\rm (S)}_H$ is a sufficient, but not necessary, condition for being embeddable; $v<v_H^{\rm (S)}$ is a necessary, but not sufficient, condition for being embeddable and $\mathfrak{s}<\mathfrak{s}^{\rm (S)}$ is a necessary and sufficient condition for being embeddable in $\mathbb{E}^3$. It would be interesting to test the generality of these results in other families of BH solutions. In this respect, we have verified the threshold value of the sphericity $\mathfrak{s}^{\rm (S)}$ still holds for the Kerr-Newman~\cite{Newman:1965my} and Kerr-Sen BH solutions~\cite{Sen:1992ua}. 

Another possible avenue for further work would be to investigate the horizon geometry within the framework of the
	``analytic effective model" for hairy BHs recently proposed in~\cite{Brihaye:2018woc}. 
In this model,
the horizon quantities (such as horizon area, Hawking temperature and horizon angular velocity)
of the hairy BH are well approximated by those of a Kerr BH, but
with the replacements
$(M,J)\to (M_H,J_H)$.
As discussed in~\cite{Herdeiro:2017phl,Brihaye:2018woc}, this model works well in the neighbourhood of the existence line
wherein the hairy BHs reduce to the vacuum Kerr solution.
Our preliminary results suggest that
the ``analytic effective model"
 holds also at the level of the horizon geometry.
That is,
in a region close to the existence line, the
horizon 
geometry~\eqref{InducedMetricKBHsSH} is well approximated by the following expressions
\begin{eqnarray}
\label{sup1}
&&
g_{\theta \theta}=r_H^2 e^{2F_1(r_H,\theta)}
=2M_H^2 \left(1+\sqrt{1-j_H^2}-\frac{j_H^2}{2}\sin^2\theta \right),~~
\\
&&
\nonumber
g_{\varphi \varphi}=r_H^2 e^{2F_2(r_H,\theta)}\sin^2\theta
=\frac{4M_H^2 \left(1+\sqrt{1-j_H^2} \right)\sin^2\theta }{1+\sqrt{1-j_H^2}
+(1-\sqrt{1-j_H^2})\cos^2\theta} \ .
\end{eqnarray}
This approximation holds within the same errors as those  
reported in~\cite{Herdeiro:2017phl,Brihaye:2018woc} 
for various physical quantities.

\newpage

\section*{Acknowledgements}

J. D. was supported by the FCT grant SFRH/BD/130784/2017. C. H. and E.R. acknowledge funding from the FCT-IF programme.  This work was also supported by the European  Union's  Horizon  2020  research  and  innovation  programme  under  the H2020-MSCA-RISE-2015 Grant No.   StronGrHEP-690904, the H2020-MSCA-RISE-2017 Grant No. FunFiCO-777740  and  by  the  CIDMA  project UID/MAT/04106/2013. The authors  would also  like  to  acknowledge networking support by the COST Action GWverse CA16104.



\bibliography{BibliographyKBHsSH}{}
\bibliographystyle{ieeetr}

\end{document}